\documentclass[%
 reprint,
nofootinbib,
 amsmath,amssymb,
 aps,
 pra,
 eqsecnum  
]{revtex4-2}

\usepackage{graphicx}
\usepackage{dcolumn}
\usepackage{bm}
\usepackage{hyperref}

\hypersetup{colorlinks=true,linkcolor=blue,urlcolor=blue,citecolor=blue} 

\usepackage{hypernat} 

\begin{document}

\preprint{APS/123-QED}

\title{Quantum-Classical Dynamical Brackets}

\author{M. Amin}
 \email{m.amin@uleth.ca}
\author{M.A. Walton}%
 \email{walton@uleth.ca}
\affiliation{%
    Department of Physics and Astronomy, University of Lethbridge, Lethbridge, Alberta, Canada, T1K 3M4
}%




\date{\today}

\begin{abstract}
    We study the problem of constructing a general hybrid quantum-classical bracket from a partial classical limit of a full quantum bracket.  Introducing a hybrid composition product, we show that such a bracket is the commutator of that product.  From this we see that the hybrid bracket will obey the Jacobi identity and the Leibniz rule provided the composition product is associative.  This suggests that the set of hybrid variables belonging to an associative subalgebra with the composition product will have consistent quantum-classical dynamics.  This restricts the class of allowed quantum-classical interaction Hamiltonians.  Furthermore, we show that pure quantum or classical variables can interact in a consistent framework, unaffected by no-go theorems in the literature or the restrictions for hybrid variables.  In the proposed scheme, quantum backreaction appears as quantum-dependent terms in the classical equations of motion.
\end{abstract}

\maketitle


\section{Introduction}

A consistent framework for the dynamics of interacting quantum and classical systems is desired for applications ranging from chemical physics to inflationary cosmology (see \cite{boucher_semiclassical_1988,prezhdo_mixing_1997} and references therein for a list of applications).  Another motivation for investigating quantum-classical hybrid dynamics is that it might shed light on the problem of quantum measurement.

Schemes for combining quantum and classical mechanics have been proposed in the literature.  For a review of the various approaches and their shortcomings, see \cite{barcelo_hybrid_2012}.  In this paper we are concerned with the approach that attempts to construct a dynamical bracket in the canonical formulation.  Taking this route allows us a  transparent, albeit abstract, analysis of the proposed dynamics and a concrete comparison with the core properties of quantum and classical dynamics.

Previous attempts of this sort have been proposed in \cite{aleksandrov_statistical_1981, gerasimenko_1982, boucher_semiclassical_1988, anderson_quantum_1995, prezhdo_quantum-classical_2006,PhysRevA.85.052109}.  They have been criticized in a series of no-go theorems presented in \cite{salcedo_absence_1996, caro_impediments_1999, sahoo_mixing_2004, salcedo_statistical_2012, gil_canonical_2017}. The suggestions fail to satisfy two crucial properties of dynamical brackets: (i)~the Jacobi identity and (ii)~the Leibniz rule.

Impediments to a consistent mechanics for a hybrid quantum-classical system have been known for quite some time now.  These no-go results are surprising, especially since part-classical part-quantum systems appear to be common.

An exploration of more general frameworks is warranted. Here we consider one such possibility, by introducing a nontrivial composition product for hybrid quantum-classical observables.  Deriving the hybrid bracket as a partial classical limit of a full quantum bracket, it takes the form of a commutator of a certain product which we call the \textit{hybrid composition product}.  The consistency of the bracket then requires that the hybrid composition product is associative.  The associativity of the product ensures the Jacobi identity and the Leibniz rule for the bracket.  It is important to note that the Leibniz rule is satisfied only with respect to the hybrid composition product.

This clear condition then allows us to investigate the problem on concrete grounds.  The hybrid bracket and its underlying composition product are derived through a partial classical limit of quantum mechanics in phase space.  The outcome of this limit depends on the quantization scheme used prior to taking the limit.  We find that the composition product resulting from familiar quantizations is not generally associative for all possible hybrid variables and thus the resulting hybrid bracket is not consistent.  This can be viewed as a no-go theorem.  However, using the associativity condition, we can explore possible ways forward.

Using the consistency condition in its alternative form, we see that hybrid variables that form an associative subalgebra with the hybrid composition product will automatically have consistent dynamics.  The associativity condition helps us find exactly which variables are admissible into the theory.  Since interaction Hamiltonians are necessarily hybrid, the condition then dictates the kind of interactions allowed between quantum and classical systems.

Furthermore, we show that pure quantum and pure classical variables can interact consistently without restriction.  The restriction on hybrid variables is relevant only if one is interested in the dynamics of hybrid variables.  For pure variables, interacting through a hybrid Hamiltonian, the dynamics is consistent.  On the quantum side, this is nothing new; it is the familiar quantum evolution on a classical background.  On the classical side, however, this gives rise to quantum backreaction on classical variables.  

In obtaining these results, no extra assumptions are made.  The hybrid setup, along with its allowed set of hybrid variables, comes naturally out of the partial classical limit.  While hybrid dynamics is not generally consistent for all hybrid variables, quantum mechanics and its partial classical limit already contain a large class of hybrid variables that have consistent dynamics.  This class of variables can be found using the methods described in this paper.

We derive the hybrid bracket from a partial application of the classical limit. The phase-space formulation of quantum mechanics is used.  It provides a general lucid transition from quantum to classical, avoiding the possible confusions of defining such transitions in operator quantum mechanics \cite{ballentine_quantum_1998}.  Of course, the phase-space formulation is algebraically equivalent to the operator formulation, so it is straightforward to translate the bracket obtained into the more familiar language of operators.

This paper is arranged as follows.  In Sec.~\ref{sec:taste} we outline the main ideas of the paper and the form of a general quantum-classical bracket.  The derivation of the bracket is shown in Sec.~\ref{sec:derivation} after a brief review of quantum mechanics in phase space.  We discuss the consistency of the dynamics of interacting quantum and classical variables and backreaction in Sec.~\ref{sec:interaction}.  In Sec.~\ref{sec:JL} we detail the condition for the hybrid bracket to obey the Jacobi identity and the Leibniz rule.  The allowed classes of hybrid variables are discussed in Sec.~\ref{sec:subalg} along with some examples of such classes.  Finally, the dynamics of hybrid variables in the Heisenberg and Schrödinger pictures is presented in Sec.~\ref{sec:dynamics}.


\section{The hybrid bracket}\label{sec:taste}

A consistent dynamical framework in the canonical formulation requires the existence of a dynamical bracket: an antisymmetric bilinear bracket of dynamical variables that obeys the Jacobi identity and the Leibniz rule.\footnote{Bilinearity is guaranteed if the bracket is both linear and antisymmetric, as in the case for this work.  However, we use the term ``bilinear'' since not all hybrid brackets proposed in the literature are antisymmetric, e.g., Anderson's bracket~\cite{anderson_quantum_1995}.}  In classical mechanics the Poisson bracket is used.  In quantum mechanics the dynamical bracket is the commutator.  It is well known that the loss of any of these four properties (antisymmetry, bilinearity, Jacobi, and Leibniz) leads to the breakdown of the dynamical framework (see~\cite{caro_impediments_1999} for a discussion of the necessity of these properties).  We expect the canonical quantum-classical (hybrid) dynamics to possess the same properties.

In~\cite{aleksandrov_statistical_1981, gerasimenko_1982, boucher_semiclassical_1988}, the bracket
\begin{align}\label{eq:AGBT}
    \frac{1}{i\hbar} [\hat{A},\hat{B}] + \frac{1}{2}\left(\{\hat{A},\hat{B}\}-\{\hat{B},\hat{A}\}\right)
\end{align}
was proposed as a hybridization of the quantum and classical brackets.  Here $\hat{A}$ and $\hat{B}$ are hybrid variables.  The bracket~\eqref{eq:AGBT} was derived by taking the classical limit of only a part of the system.  While antisymmetric and bilinear, it does not generally satisfy the Jacobi identity or the Leibniz rule.  This negative result was discussed in the no-go theorems of~\cite{salcedo_absence_1996, caro_impediments_1999, sahoo_mixing_2004, salcedo_statistical_2012, gil_canonical_2017}, forbidding a consistent framework for quantum-classical dynamics.

In this paper we apply the partial classical limit in a more general context as opposed to the special case used in the literature.  In doing so, a single mathematical condition emerges as responsible for the Jacobi and Leibniz properties.  This condition then puts the aforementioned no-go theorems in a more general context and provides a language in which they can be further analyzed.  We explore the possibilities for hybrid quantum-classical dynamics in light of this realization.

To arrive at the consistency condition, we first observe some general properties of both quantum and classical mechanics.  Dynamical variables can be combined to produce new ones through a bilinear and associative binary operation.  We call this operation a \textit{composition product}.  The composition product of quantum mechanics is noncommutative, while that of classical mechanics is commutative.  Dynamical brackets satisfy the conditions of antisymmetry, bilinearity, and the Jacobi identity
\begin{align}
    [\hat{A},[\hat{B},\hat{C}]] &= [[\hat{A},\hat{B}],\hat{C}] + [\hat{B},[\hat{A},\hat{C}]]~,\\
    \{f,\{g,h\}\} &= \{\{f,g\},h\} + \{g,\{f,h\}\}~,
\end{align}
and they obey the Leibniz rule \textit{with respect to the composition product}
\begin{align}
    [\hat{A},\hat{B}\hat{C}] &= [\hat{A},\hat{B}]\hat{C} + \hat{B}[\hat{A},\hat{C}]~,\\
    \{f,gh\} &= \{f,g\}h + g\{f,h\}~.
\end{align}
Here $\hat{A}$, $\hat{B}$, and $\hat{C}$ are quantum operators on Hilbert space, and $f$, $g$, and $h$ are classical functions on phase space.

Realizing this abstract structure of composition products and dynamical brackets provides insight into the problem of constructing a consistent dynamical framework.  In particular, we need to specify the composition product with respect to which a dynamical bracket obeys the Leibniz rule.

In quantum mechanics, the noncommutativity of the composition product allows the commutator of this product to be a good candidate for a dynamical bracket.  Indeed, the noncommutativity and bilinearity of the quantum product lead to the antisymmetry and bilinearity of the commutator.  Its associativity ensures that the commutator of the quantum product obeys the Jacobi identity and the Leibniz rule with respect to that same product.  We see that the composition product itself gives rise to a dynamical bracket.

On the other hand, the composition product of classical mechanics is commutative.  Constructing a dynamical bracket for it follows a different path.  We can write the Poisson bracket as a commutator of a bidifferential operator.  For example,
\begin{align}
    \{f,g\} = f \overleftarrow{\partial}_q \overrightarrow{\partial}_p\, g - g \overleftarrow{\partial}_q \overrightarrow{\partial}_p\, f~.
\end{align}
Here $q$ and $p$ are the canonical phase-space conjugates and we use the standard notation $f \overleftarrow{\partial} := \partial f$ and $\overrightarrow{\partial} f := \partial f$.
The noncommutativity and bilinearity of $\overleftarrow{\partial}_q \overrightarrow{\partial}_p$ lead to the antisymmetry and bilinearity of the Poisson bracket.  However, $\overleftarrow{\partial}_q \overrightarrow{\partial}_p$ is not associative.  Only by asserting that the familiar pointwise multiplication is the classical composition product can we find that the Poisson bracket satisfies the Jacobi identity and the Leibniz rule with respect to that composition product.

We can then approach hybrid dynamics using the lens of composition products and dynamical brackets.  As we show in the coming sections, the composition product of hybrid variables, like that of pure quantum variables, is noncommutative.  We denote it by $\circledast$ and call it the asterisk product.  Thus, writing the hybrid bracket as a commutator of this new composition product
\begin{align}\label{eq:taste}
    \{\![\hat{A},\hat{B}]\!\} = \frac{1}{i\hbar} \left( \hat{A} \circledast \hat{B} - \hat{B} \circledast \hat{A} \right)~,
\end{align}
we see immediately that we must require $\circledast$ to be associative.  The associativity of the $\circledast$-product\footnote{In this paper, we use the term ``product'' to refer to the binary operation itself: $\circledast$ is the product, not $f \circledast g$.} implies the Jacobi identity for its commutator~\eqref{eq:taste} and provides the Leibniz rule as
\begin{align}
    \{\![\hat{A},\hat{B} \circledast \hat{C}]\!\} = \{\![\hat{A},\hat{B}]\!\} \circledast \hat{C} + \hat{B} \circledast \{\![\hat{A},\hat{C}]\!\}~.
\end{align}
Together with noncommutativity and bilinearity, the associativity of $\circledast$ is the condition that guarantees the Jacobi and Leibniz properties for the hybrid bracket.

We derive the bracket~\eqref{eq:taste} in the next section by taking a partial classical limit of the system.  Also used are the natural reduction requirements~\cite{caro_impediments_1999}:
\begin{subequations}\label{eq:red-taste}
\begin{align}
    \{\![\text{quantum}\,,\text{classical}]\!\} &= 0~,\\
    \{\![\text{quantum}\,,\text{any}]\!\} &= \frac{1}{i\hbar} [\text{quantum}\,,\text{any}]~,\\
    \{\![\text{classical}\,,\text{any}]\!\} &= \{\text{classical}\,,\text{any}\}~.
\end{align}
\end{subequations}
For now, we simply state that 
\begin{align}\label{eq:ast-taste}
    \circledast = 1 + \frac{i\hbar}{2}(\mathcal{P}+\sigma)
\end{align}
where $\mathcal{P}$ is the Poisson bracket
\begin{align}
    f \, \mathcal{P} \, g := f \left( \overleftarrow{\partial}_{q} \overrightarrow{\partial}_{p} - \overleftarrow{\partial}_{p} \overrightarrow{\partial}_{q} \right) g = \{f,g\}~,
\end{align}
and $\sigma$ is symmetric ($g\,\sigma\, f=f\,\sigma\, g$, or $\sigma^t=\sigma$).  The $\sigma$-product involved in the hybrid composition product $\circledast$ and bracket $\{\![\cdot\,,\cdot]\!\}$ is connected to the partial classical limit in the next section.  In general, $\sigma$ defines the $\circledast$-product and, consequently, the hybrid bracket.

Using~\eqref{eq:ast-taste} in~\eqref{eq:taste}, the hybrid bracket becomes
\begin{align}\label{eq:taste2}
    \frac{1}{i\hbar} [\hat{A},\hat{B}] + \frac{1}{2}\left( \{\hat{A},\hat{B}\}-\{\hat{B},\hat{A}\} + \hat{A}\,\sigma\,\hat{B} - \hat{B}\,\sigma\,\hat{A} \right)~.
\end{align}
The bracket~\eqref{eq:AGBT} is a special case ($\sigma=0$) of this general result.  Note that, regardless of the commutativity of $\mathcal{P}$ or $\sigma$ as binary operations, $\hat{A}$ and $\hat{B}$ are noncommutative in the general case.

The bracket~\eqref{eq:taste2} is antisymmetric, bilinear, and obeys the reduction requirements~\eqref{eq:red-taste}.  For this bracket to obey the Jacobi identity and the Leibniz rule, the $\circledast$-product~\eqref{eq:ast-taste} must be associative, which then places a condition on $\sigma$.  As will be seen in Sec.~\ref{sec:JL}, straightforward constructions of $\sigma$ cannot satisfy that condition.  This can be seen as a generalization of the no-go theorem to certain hybrid brackets.

The analysis presented thus far connects the consistency of the framework of hybrid dynamics to the associativity of the $\circledast$-product.  This connection suggests the possibility of circumventing previous no-go theorems.  To that end, one path to explore is the construction of a nontrivial $\circledast$-product that is associative for general hybrid variables.  Another possibility is to restrict the allowable hybrid variables to those that form an associative subalgebra with a given $\circledast$-product.  The latter will be discussed in Sec.~\ref{sec:subalg}.

\section{Derivation from the partial classical limit}\label{sec:derivation}

In an attempt to construct hybrid mechanics, one can start with the assumption that quantum mechanics underlies classical systems.  In this setup, the dynamical variables are subdivided into $Q$ and $C$ sectors.  We then proceed to take the classical limit of the $C$ sector only, obtaining a general form of the hybrid dynamical bracket.

Defining the classical limit is not straightforward.  Various ways of obtaining such a limit in the standard operator formulation have proven difficult \cite{ballentine_quantum_1998}, which prevents them from providing a procedure general enough for our purposes. The phase-space formulation of quantum mechanics, on the other hand, provides a sufficiently general and unambiguous correspondence between the algebras of classical and quantum mechanics in the limit of $\hbar \to 0$. 

Here we present a brief overview of the concepts of phase-space quantum mechanics (also known as deformation quantization) that we use to study quantum-classical brackets.  The interested reader is referred to \cite{takahashi_distribution_1989,cohen_generalized_1966,cohen_weyl_2013,lee_theory_1995,BalazsJennings84,blaszak_phase_2011,zachos_quantum_2005,bayen_deformation_1978-I,bayen_deformation_1978-II} for more extensive treatments, as well as to the more pedagogical reviews \cite{ballentine_quantum_1998,hirshfeld_deformation_2002,hancock_quantum_2004,case_wigner_2008}.

The phase-space formulation is generated from operator quantum mechanics by mapping Hilbert space operators to phase-space functions:
\begin{align}\label{eq:map}
    \hat{A}(\hat{q},\hat{p}) &\to A_R(q,p)~.
\end{align}
Here $q$ and $p$ stand for any number of degrees of freedom and their conjugate momenta.  The map, or \textit{realization}, $R$ is sometimes called a dequantization and it is not unique. For example, the quantization $R^{-1}$ entails an operator ordering prescription, and so the map $R$ depends on which one is chosen.      

Phase-space quantum mechanics comes with a noncommutative, bilinear, and associative $\star$-product (star product) mimicking (homomorphic to) the noncommutative product of operators 
\begin{align}\label{deqn}
    \hat{A}\hat{B} &\to A_R \star_R B_R~.
\end{align}
The quantum bracket is the commutator of the $\star$-product divided by $i\hbar$
\begin{align}
    [\![\cdot\,,\cdot]\!] = \frac{1}{i\hbar} (\star - \star^t)~.
\end{align}
Here $\hbar$ is the reduced Planck constant and the superscript $t$ stands for transpose: $A \star^t B = B \star A$.

A $\star$-product is a ``deformation'' of the commutative, bilinear, and associative pointwise product (ordinary multiplication) of functions, with $i\hbar$ as a deformation parameter.  These deformations should obey the general classical correspondence relations 
\begin{align}
    \star &= \sum_{n=0}^\infty (i\hbar)^n \mathcal{G}_n = 1 + \mathcal{O}(\hbar)~,\label{eq:corr1}\\
    [\![\cdot\,,\cdot]\!] &= \sum_{n=1}^\infty (i\hbar)^{n-1} \left(\mathcal{G}_n - \mathcal{G}_n^t\right) = \mathcal{P} + \mathcal{O}(\hbar)~,\label{eq:corr2}
\end{align}
where $\mathcal{G}_n$ are noncommutative binary operations (products) and $\mathcal{P}$ is the Poisson bracket.  Following the nonuniqueness of the operator-to-function map (\ref{eq:map}), the deformation is not unique.  Since all $\star$-products have their zeroth term as defined above, the particular form of the (generally noncommutative) $\hbar$-dependent terms will reflect the difference between $\star$-products.

A class of representations can be obtained from  different quantization maps, specifying different operator-ordering recipes.  The most famous $\star$-product on phase space is the one based on the Wigner transform, reflecting the Weyl operator ordering,
\begin{align}
    \star_W &= \text{e}^{\frac{i\hbar}{2} \mathcal{P}}~.
\end{align}
Other examples include star products based on the standard, antistandard, normal, antinormal, and Born-Jordan orderings.  
It is possible to relate two realizations via a transition operator~\cite{bayen_deformation_1978-I,bayen_deformation_1978-II}.

This concludes our very brief summary.  Let us now take the partial classical limit and derive the general hybrid bracket.  We subdivide the degrees of freedom and their conjugate momenta into two sectors, referred to as $Q$ and $C$.  Any dynamical variable $u$ can be put in the form $u = \sum u_Q u_C$.  The sum is over any number of variables and the subscripts refer to variables belonging to the $Q$ or $C$ sectors.  Since we will be dealing with bilinear products, we will restrict our discussion to hybrid variables of the form $u = u_Qu_C$, omitting the sum; the results are still general.

The $\star$-product of the full quantum system can be factored:
\begin{align}\label{eq:star-full}
    \star_{\textup{full}} = \star_Q \star_C = \star_C \star_Q~.
\end{align}
Here the subscripts imply that each of the $\star$-products acts only on its respective sector, 
\begin{align}
    u \star_{\textup{full}} v = u_Qu_C \star_Q\star_C v_Qv_C = (u_Q\star_Qv_Q) (u_C\star_Cv_C)~.
\end{align}
This ensures that variables belonging to two different sectors commute.  It should be understood that $u_Q$ and $u_C$ are pure $Q$ and pure $C$ variables, respectively.

Note that, since the two sectors are independent, $\star_Q$ need not be realized in the same way as $\star_C$.  In the following, only the $C$ sector realization is relevant; the freedom of choosing any quantization scheme on the $Q$ sector is maintained.

The partial classical limit then is the process of taking $\hbar\to 0$ only in the $C$ sector of the system. The result should apply to systems whose $C$ sector has a scale of action, constructed from its physical parameters, that is large compared to $\hbar$.   

From the correspondence relations (\ref{eq:corr1}) and (\ref{eq:corr2}) with the factorized $\star$-product~\eqref{eq:star-full}, we have
\begin{align}
    \star_{full} &= \star_Q\left[1+i\hbar_C \mathcal{G_C}_1 + \mathcal{O}(\hbar_C^2)\right]~,\label{eq:starfull}\\
    \begin{split}
    [\![\cdot\,,\cdot]\!]_{full} &= \frac{1}{i\hbar} \left(\star_{full} - \star_{full}^t\right)\label{eq:qbfull}\\
    &= [\![\cdot\,,\cdot]\!]_Q + \frac{\hbar_C}{\hbar} \left(\star_Q\mathcal{G_C}_1 - \star_Q^t\mathcal{G_C}_1^t\right) + \mathcal{O}\left(\frac{\hbar_C^2}{\hbar}\right)~,
    \end{split}
\end{align}
where $\hbar_C$ is Planck's constant on the $C$ sector.  Taking the partial classical limit $\hbar_C^2/\hbar\to0$ in (\ref{eq:qbfull}) gives the hybrid bracket
\begin{align}\label{eq:hybrid0}
    \{\![\cdot\,,\cdot]\!\} = [\![\cdot\,,\cdot]\!] + \frac{\hbar_C}{\hbar} (\star\mathcal{G}-\star^t\mathcal{G}^t)~.
\end{align}
Here we suppress the subscripts for simplicity: $\star := \star_Q$ and $\mathcal{G} := \mathcal{G_C}_1$.  From now on, $\star$ should be understood as acting only on $Q$ sector variables and $\mathcal{G}$ on $C$ sector variables.

The uniqueness, or lack thereof, of Planck's constant has been discussed in~\cite{sahoo_mixing_2004, salcedo_statistical_2012}.  Here we take Planck's constant to be unique and universal; numerically, $\hbar_C = \hbar$.  We keep the subscript in $\hbar_C$ simply as a bookkeeping device in the formal expansion of the full quantum bracket $[\![\cdot\,,\cdot]\!]_\text{full}$ [Eq.~\eqref{eq:qbfull}].  More concretely, the expansion should be of the bracket \textit{of} physical quantities.  Explicitly, taking the limit $\hbar_C\to0$ means that $\hbar$ is small compared to some characteristic scale of action $S$ constructed from physical quantities in the $C$ sector:
\begin{align}
    \frac{\hbar}{S(u_C,v_C)} \ll 1~.
\end{align}
However, since we do not include physical quantities in the formal expansion~\eqref{eq:qbfull}, we use the symbol $\hbar_C$ as the subject of the partial classical limit while keeping ``the other'' $\hbar$ on the $Q$ sector untouched.

Now that the partial classical limit has been taken and the hybrid bracket has been derived, we set $\hbar_C=\hbar$ in~\eqref{eq:hybrid0} to get
\begin{align}\label{eq:hybrid1}
    \{\![\cdot\,,\cdot]\!\} = [\![\cdot\,,\cdot]\!] + \star\,\mathcal{G}-\star^t\,\mathcal{G}^t~.
\end{align}

The use of phase-space quantum mechanics allows a simple derivation of a general quantum-classical hybrid bracket.  The expression for the hybrid bracket (\ref{eq:hybrid1}) will differ depending on the choice of representation on the $C$ sector.  If one chooses to use the Wigner representation with its associated $\star_C$-product on the $C$ sector, then one gets $\mathcal{G} = \mathcal{P}/2$ and the bracket becomes
\begin{align}\label{eq:abt}
    [\![\cdot\,,\cdot]\!] + \frac{\star+\star^t}{2}\mathcal{P}~.
\end{align}
This is the bracket proposed in \cite{aleksandrov_statistical_1981, gerasimenko_1982, boucher_semiclassical_1988}, thus making it a special case of the general bracket.

Directly from the associativity of $\star_C = 1 + i\hbar\, \mathcal{G} + \cdots$ and the correspondence relations (\ref{eq:corr1}) and (\ref{eq:corr2}), we have two important properties of $\mathcal{G}$
\begin{align}
    u\, (v\, \mathcal{G}\, w) - (uv)\, \mathcal{G}\, w + u\, \mathcal{G}\, (vw) - (u\, \mathcal{G}\, v)\, w = 0~,\label{eq:acond}\\
    \mathcal{G} - \mathcal{G}^t = \mathcal{P}~.\qquad\qquad\qquad\quad\label{eq:pcomm}
\end{align}
Equation~(\ref{eq:acond}) gives us a relation between $\mathcal{G}$ and normal multiplication, while Eq.~(\ref{eq:pcomm}) shows that the Poisson bracket is the commutator, or the antisymmetric part, of the $\mathcal{G}$-product.  We can then write $\mathcal{G}$ as a sum of antisymmetric and symmetric (commutative) parts
\begin{align}\label{eq:parts}
    \mathcal{G} = \frac{1}{2}\left(\mathcal{P} + \sigma\right)~.
\end{align}
The form of $\sigma$ differs for different realizations, while $\mathcal{P}$ is the antisymmetric part of $\mathcal{G}$ in all.  For example, $\sigma=0$ for the Wigner or the Born-Jordan-based representations.  

Using (\ref{eq:parts}), the hybrid bracket (\ref{eq:hybrid1}) can be rewritten as
\begin{align}\label{eq:hybrid2}
    \{\![\cdot\,,\cdot]\!\} = [\![\cdot\,,\cdot]\!] + 
    \frac{\star+\star^t}{2}\mathcal{P} + \frac{\star-\star^t}{2}\sigma ~,
\end{align}
giving it a somewhat symmetric form and allowing for comparison with different particular realizations of the bracket.  Using this form, we can test whether the bracket satisfies the natural assumptions discussed in \cite{caro_impediments_1999},
\begin{align}\label{eq:reduction}
    \begin{split}
    \{\![u_Q,v_C]\!\} &= 0~,\\
    \{\![u_Q,v]\!\} &= v_C[\![u_Q, v_Q]\!]~,\\
    \{\![u_C,v]\!\} &= v_Q\{u_C,v_C\}~,
    \end{split}
\end{align}
where $v=v_Qv_C$ is an arbitrary hybrid variable.  These conditions ensure that separate sectors remain separate and that hybrid dynamics reduces to pure dynamics for pure (quantum or classical) variables.

For the bracket (\ref{eq:hybrid2}), the first and last of these relations are identically satisfied, while the middle one is calculated to be
\begin{align}\label{eq:conq}
    \{\![u_Q,v]\!\} &= \big[v_C+(i\hbar/2) (1\, \sigma\, v_C) \big] [\![u_Q,v_Q]\!]~.
\end{align}
Requiring Eq.~(\ref{eq:conq}) to follow the exact equality in (\ref{eq:reduction}) imposes the condition
\begin{align}\label{eq:1sigma}
    1~\sigma~u = u~\sigma~1 = 0
\end{align}
for any variable $u$.  From (\ref{eq:parts}) we get the same condition for $\mathcal{G}$.  Now the hybrid bracket reduces to pure brackets, as desired.

There is another form of the general hybrid bracket that will prove very useful in recognizing its properties.  Defining
\begin{align}\label{eq:ast}
    \circledast := 1+i\hbar\,\mathcal{G}~,
\end{align}
the bracket becomes
\begin{align}\label{eq:hybrid3}
    \{\![\cdot\,,\cdot]\!\} = \frac{1}{i\hbar}\left(\star\circledast-\star^t\circledast^t\right)~.
\end{align}
The hybrid bracket is essentially a commutator of the $\star\circledast$-product with $\star$ acting only on $Q$ variables and $\circledast$ on $C$ variables.

The $\circledast$-product~\eqref{eq:ast} can be seen as a truncation of the $\star$-product at the first power of $\hbar$.  Stopping at the first-order truncation is motivated by two reasons: (i)~It gives the partial classical limit of the full quantum dynamical bracket~\eqref{eq:qbfull}, and (ii)~it is the only truncation compatible with the reduction requirements, particularly, the third of~\eqref{eq:reduction}.

Another useful form of the hybrid bracket can be found using Eqs.~\eqref{eq:hybrid1}, \eqref{eq:parts}, and~\eqref{eq:ast}.  It has the character of a Leibniz rule
\begin{align}\label{eq:hybrid4}
    \{\![u,v]\!\} = [\![u,v_Q]\!] \circledast v_C + v_Q \star\{u,v_C\}~.
\end{align}
This shows that pure factors of a hybrid variable $v = v_Q v_C$ can be pulled out of the hybrid bracket at the cost of $\star$- and $\circledast$-products.  From this we see immediately that a conserved pure quantum variable $v_Q$ and a pure classical variable $v_C$ can be combined to give a conserved hybrid variable $v=v_Qv_C$.

The quantum-classical bracket can be expressed in terms of operators as follows.  Consider hybrid dynamical entities of the form $\hat{A}(q_C,p_C) = \hat{A}_Q A_C(q_C,p_C)$, where $\hat{A}_Q$ is an operator on the Hilbert space of the $Q$ sector and $A_C(q_C,p_C)$ is a function on the phase space of the $C$ sector.  The hybrid bracket can be written as
\begin{align}\label{eq:hybrid5}
    \{\![\hat{A},\hat{B}]\!\} = \frac{1}{i\hbar}\left(\hat{A}\circledast\hat{B}-\hat{B}\circledast\hat{A}\right)~,
\end{align}
or equivalently
\begin{align}\label{eq:hybrid6}
    \frac{1}{i\hbar}[\hat{A},\hat{B}] + \frac{1}{2}
    \left( \{\hat{A},\hat{B}\} - \{\hat{B},\hat{A}\} + \hat{A}\sigma\hat{B} - \hat{B}\sigma\hat{A} \right)~.
\end{align}
This again offers a direct comparison with the bracket proposed in \cite{aleksandrov_statistical_1981, gerasimenko_1982, boucher_semiclassical_1988}.  Finally, the operator form of the quasi-Leibniz relation~\eqref{eq:hybrid4} is
\begin{align}\label{eq:hybrid7}
    \{\![\hat{A},\hat{B}]\!\} = [\![\hat{A},\hat{B}_Q]\!] \circledast B_C + \hat{B}_Q \{\hat{A},B_C\}~.
\end{align}
We will return to, and continue to use, the phase-space formulation.


\section{Dynamics of interacting pure variables}\label{sec:interaction}

The general hybrid bracket, in all its incarnations mentioned in the preceeding section, obeys some important consistency requirements.  It reduces to a pure bracket when one of its arguments is pure [Eq.~\eqref{eq:reduction}] and it is antisymmetric and bilinear for all variables.  Before addressing the Jacobi identity and Leibniz rule for hybrid variables, it is instructive to see what can be done with the bracket as it is, using only the properties it already satisfies.

The reduction equations~\eqref{eq:reduction} show a feature of the hybrid bracket that can be exploited immediately: When the hybrid bracket reduces to a pure one, it automatically satisfies all consistency requirements, including the Jacobi identity and Leibniz rule.  Reduction is guaranteed when there is only one independent hybrid variable considered.  This is a plausible setup for quantum and classical systems interacting through a hybrid interaction term in the Hamiltonian\footnote{The Hamiltonian can also be a piecewise function that has different forms in different regions of the phase space or time.  The point of the restriction is to not have two hybrid variables in one dynamical bracket.}.  This has the implication that only time translation can be generated by a hybrid variable; no other hybrid transformations are allowed.  Despite the restriction on hybrid variables, the implications for the familiar pure variables are significant.  In fact, they would cover a wide range of applications.

In this section we will consider the case where all dynamical variables of interest are pure (quantum or classical) except for one hybrid variable that defines the interaction Hamiltonian.  Let $\eta(q_Q,p_Q,q_C,p_C)$ be a hybrid variable.  Then, by reduction, we have
\begin{align}
    \{\![u_Q,f(\eta)]\!\} &= [\![u_Q,f(\eta)]\!]~,\\
    \{\![u_C,f(\eta)]\!\} &= \{u_C,f(\eta)\}~,
\end{align}
and from antisymmetry
\begin{align}
    \{\![f(\eta),g(\eta)]\!\} &= 0
\end{align}
for all functions of $\eta$.

Let the Hamiltonian of the full system be
\begin{align}
    H = H_Q + H_C + H_I(\eta)~,
\end{align}
split into a pure quantum term $H_Q$, a pure classical term $H_C$ and an interaction term $H_I(\eta)$.  For any function $f_Q$ of pure quantum variables, the analog of the Heisenberg equation of motion is then
\begin{align}
    \begin{split}
    \frac{df_Q}{dt} &= \{\![f_Q,H]\!\} + \frac{\partial f_Q}{\partial t}\\
    &= [\![f_Q,H_Q+H_I]\!] + \frac{\partial f_Q}{\partial t}~,
    \end{split}
\end{align}
the expected quantum equation.  The interaction term $H_I$ contains classical variables that are no different from the familiar external potentials in textbook quantum mechanics.  From the point of view of the $Q$ sector, $H_I(\eta)$ can be regarded as a time-dependent potential where the explicit time dependence is that of the classical variables involved in $\eta$.

On the classical side, we get a similar equation:
\begin{align}\label{eq:c-dot}
    \begin{split}
    \frac{df_C}{dt} &= \{\![f_C,H]\!\} + \frac{\partial f_C}{\partial t}\\
        &= \{f_C,H_C+H_I\} + \frac{\partial f_C}{\partial t}~.
    \end{split}
\end{align}
This time, the interaction term $H_I(\eta)$ encodes quantum backreaction on classical variables.  This is due to the presence of quantum variables in $\eta$, which evolve quantum mechanically.

The interaction term, as a whole, has a hybrid time evolution
\begin{align}\label{eq:HI}
    \begin{split}
    \frac{dH_I}{dt} &= \{\![H_I,H]\!\} + \frac{\partial H_I}{\partial t}\\
        &= [\![H_I,H_Q]\!] + \{H_I,H_C\} + \frac{\partial H_I}{\partial t}~.
    \end{split}
\end{align}
In general, any function of $\eta$ will evolve in time according to
\begin{align}\label{eq:eta}
        \frac{df(\eta)}{dt} 
        = [\![f(\eta),H_Q]\!] + \{f(\eta),H_C\} + \frac{\partial f(\eta)}{\partial t}~.
\end{align}

We see that, despite the restriction of allowing only one hybrid variable (the interaction) into the theory, interesting results can still be found.  In~\cite{amin2021illustration} we provide an explicit example of the dynamics of pure variables interacting via a quantum-classical harmonic oscillator.  Backreaction is exhibited in the evolution of a nonvanishing commutator for classical variables.  Importantly, it was shown that while canonical quantum or classical relations in general are not preserved in a quantum-classical interaction, the \textit{hybrid} canonical relations are.

\section{Jacobi identity and Leibniz rule for hybrid variables}\label{sec:JL}

Now we turn to the issue of satisfying the Jacobi identity and the Leibniz rule for hybrid variables.  Examining the form~\eqref{eq:hybrid3} of the bracket, we can see that if the hybrid product $\star\circledast$ is associative, then it can be treated as a composition product for hybrid variables that automatically renders the bracket Leibniz and Jacobi compliant.

Since $\star$ is already associative, the condition becomes that $\circledast$ must be associative, acting only on classical variables.  This is clear from an explicit expansion of the Jacobi terms of the hybrid bracket
\begin{align}
    &\{\![\{\![u,v]\!\},w]\!\} + \{\![\{\![v,w]\!\},u]\!\} + \{\![\{\![w,u]\!\},v]\!\} \nonumber\\
    \begin{split}
        & = u_Q \star v_Q \star w_Q \left[ (u_C \circledast v_C) \circledast w_C - u_C \circledast (v_C \circledast w_C) \right]\\
    &\quad+\text{ cyclic permutations of }(u,v,w)\\
    &\quad+\text{ anticyclic permutations of }(u,v,w) = 0~.
    \end{split}
\end{align}
Since all quantum variables are arbitrary, cyclic and anticyclic permutations of $u_Q \star v_Q \star w_Q$ are in general independent and nonvanishing.  Then the condition for the bracket $\{\![\cdot\,,\cdot]\!\}$ to obey the Jacobi identity is for $\circledast$ to be associative:
\begin{align}\label{eq:asast}
	(u_C \circledast v_C) \circledast w_C = u_C \circledast (v_C \circledast w_C)~.
\end{align}
The associativity of $\star_C$, leading to Eq.~(\ref{eq:acond}), implies directly that, for $\circledast$ to be associative, $\mathcal{G}$ must be associative as well,
\begin{align}\label{eq:asg}
    (u_C\, \mathcal{G}\, v_C)\, \mathcal{G}\, w_C = u_C\, \mathcal{G}\, (v_C\, \mathcal{G}\, w_C)~.
\end{align}

On the other hand, the associativity of the $\star\circledast$-product makes it a candidate for a hybrid composition product.  The use of $\star\circledast$ as a composition product for hybrid variables automatically guarantees the Leibniz rule as
\begin{align}\label{eq:saleib}
	\{\![u,v \star\circledast~w]\!\} = \{\![u,v]\!\} \star\circledast~w + v \star\circledast~\{\![u,w]\!\}~,
\end{align}
or in operator form
\begin{align}\label{eq:saleib-operator}
    \{\![\hat{A},\hat{B} \circledast\hat{C}]\!\} = \{\![\hat{A},\hat{B}]\!\} \circledast\hat{C} + \hat{B} \circledast\{\![\hat{A},\hat{C}]\!\}~,
\end{align}
as opposed to the, at first glance, more intuitive
\begin{align}\label{eq:saleib-wrong}
    \{\![\hat{A},\hat{B} \, \hat{C}]\!\} = \{\![\hat{A},\hat{B}]\!\} \, \hat{C} + \hat{B} \, \{\![\hat{A},\hat{C}]\!\}~.
\end{align}

To be clear, in the phase-space representation, two hybrid variables $u=u_Qu_C$ and $v=v_Qv_C$ are composed as
\begin{align}
    u \star\circledast\, v = (u_Q\star v_Q)\,(u_C\circledast v_C)~.
\end{align}
Equivalently, in the operator representation, hybrid variables $\hat{A}=\hat{A}_Q A_C$ and $\hat{B}=\hat{B}_Q B_C$ are composed as
\begin{align}
    \hat{A}\circledast\hat{B} = (\hat{A}_Q\hat{B}_Q) (A_C \circledast B_C)~.
\end{align}
The $\circledast$-product acts only on the classical part of hybrid variables.

The realization that an associative $\star\circledast$-product is the natural composition product for hybrid variables puts previous no-go theorems in a larger context.  The problem now is finding a suitable composition product.  Indeed, it is the Leibniz rule that is at the heart of the counterexample in \cite{caro_impediments_1999} (proving the inconsistency of bracket (\ref{eq:abt}) as presented in \cite{aleksandrov_statistical_1981, gerasimenko_1982, boucher_semiclassical_1988}), as well as the no-go theorem in \cite{salcedo_statistical_2012}.  The Leibniz rule should take the form~\eqref{eq:saleib} or~\eqref{eq:saleib-operator} instead of~\eqref{eq:saleib-wrong}.  Equipped with the knowledge that hybrid variables need a suitable composition product, we can view the consistency problem in a different light.

The $\star\circledast$-product introduces an extra, parallel way of composing variables alongside the more familiar and physically understood $\star$-product.  It arises naturally from the construction as shown in the previous discussion.  It gives hybrid dynamics the same algebraic structure used for quantum mechanics: a noncommutative, bilinear, and (desirably) associative product that acts as a composition product and whose commutator is the dynamical bracket.

Despite reducing to $\star$ for pure quantum variables ($u_Q \star\circledast v_Q = u_Q \star v_Q$), it deviates from the familiar multiplication for pure classical variables
\begin{align}
    u_C \star\circledast v_C = u_C \circledast v_C = u_C \, v_C + i\hbar\,u_C \, \mathcal{G} \, v_C~.
\end{align}
This deviation, however, is additive, small (proportional to $\hbar$), and, importantly, does not affect classical dynamics since it does not change the Poisson bracket.

Classical mechanics too can mimic the algebraic structure of quantum mechanics provided there exists a  noncommutative, bilinear, and associative $\circledast$ acting as a composition product (with a small deviation) whose commutator builds the dynamical bracket
\begin{align}
    \{\cdot\,,\cdot\} &= \frac{1}{i\hbar}(\circledast - \circledast^t)~.
\end{align}
This similarity to quantum mechanics is not surprising.  The $\circledast$-product is a truncation of the series expansion of the $\star$-product with the condition that it is itself associative.  In that sense, the $\circledast$-product can be regarded as a semiclassical approximation of the full quantum $\star$-product.

So what is a suitable $\circledast$-product?  From Eqs.~\eqref{eq:parts} and~\eqref{eq:ast} we see that the symmetric product $\sigma$ is the part that differentiates one $\circledast$-product from another
\begin{align}\label{eq:ast-pc}
    \circledast = 1 + \frac{i\hbar}{2}\left(\mathcal{P} + \sigma\right)~.
\end{align}
Along with the condition~\eqref{eq:1sigma} that $1 \sigma u=0$, from Eqs.~\eqref{eq:parts} and~\eqref{eq:asg} we derive another condition on $\sigma$:
\begin{align}\label{eq:sigma-cond}
\begin{split}
    &(u\,\mathcal{P}\,v)\,\mathcal{P}\,w - u\,\mathcal{P}\,(v\,\mathcal{P}\,w)
    +(u\,\sigma\, v)\,\sigma\, w - u\,\sigma\, (v\,\sigma\, w)\\
    +&(u\,\mathcal{P}\,v)\,\sigma\, w - u\,\mathcal{P}\,(v\,\sigma\, w)
    +(u\,\sigma\, v)\,\mathcal{P}\,w - u\,\sigma\, (v\,\mathcal{P}\,w) \\
    =&~0~.
\end{split}
\end{align}
We see immediately that the Wigner $\star$-product ($\sigma=0$) does not satisfy this condition.  Are there other $\sigma$'s that do?

As discussed in Sec.~\ref{sec:derivation}, $\sigma$ is related to the phase-space realization, or quantization scheme, on the $C$ sector prior to the partial classical limit.  For the familiar schemes such as the Weyl, Born-Jordan, standard, antistandard, normal and antinormal ordering or the Husimi realization, $\sigma$ is first-order in the derivatives $\partial_{q_C}$ and $\partial_{p_C}$.  A general first order symmetric bidifferential operator is given by
\begin{align}\label{eq:sigma-abc}
\begin{split}
    \sigma &= a~\overleftarrow{\partial}_{q_C}\overrightarrow{\partial}_{q_C} + b~\overleftarrow{\partial}_{p_C}\overrightarrow{\partial}_{p_C} \\
    &\qquad\qquad+ c \left( \overleftarrow{\partial}_{q_C}\overrightarrow{\partial}_{p_C} + \overleftarrow{\partial}_{p_C}\overrightarrow{\partial}_{q_C} \right)~.
\end{split}
\end{align}
The constants $(a,b,c)$ correspond to different schemes.  For example, Weyl and Born-Jordan orderings are reflected by $(0,0,0)$, standard and antistandard by $(0,0,\pm 1)$, and Husimi by $(1,1,0)$.  Different values of $(a,b,c)$ define different $\sigma$'s and thus different $\circledast$'s.

We find that no choice of $(a,b,c)$ can give rise to an associative $\circledast$-product for general variables.  This is a generalization of the previous no-go theorems: The class of $\circledast$-products~\eqref{eq:ast-pc} defined by~\eqref{eq:sigma-abc} is not associative for general phase-space functions.  This means that if we insist that all possible phase-space functions should be admitted into hybrid variables, then this class of composition products cannot lead to consistent mechanics for hybrid variables.

It is not obvious, however, that the desired product cannot be constructed through a more complex setup.  The condition~\eqref{eq:sigma-cond} excludes a large class of products, which might suggest that a nontrivial framework needs to be constructed.  With the condition for consistency clearly stated in mathematical terms, we leave the investigation of this problem to future work.

Nonetheless, the realization that the consistency of hybrid dynamics hinges on the associativity of the $\circledast$-product suggests another path to explore.  That is the topic of the next section.

\section{An associative subalgebra}\label{sec:subalg}

Casting the consistency condition in terms of the associativity of the composition product $\circledast$ gives us a useful guide: \textit{dynamics is consistent for hybrid variables that form an associative subalgebra with $\circledast$}.  In other words, if we restrict our attention to a special class of hybrid variables, the hybrid bracket will obey the Jacobi identity and the Leibniz rule.  The associativity condition is a clear filter that allows to find and test hybrid variables that can be admitted into the theory.  We will see that the class of allowed hybrid variables covers a wide range of interesting cases.

As mentioned in the previous sections, the $\circledast$-product acts only on the classical part of hybrid variables.  Thus the restriction we impose produces hybrid variables of the form
\begin{align}
	f(q_Q,p_Q,q_C,p_C) = g(q_Q,p_Q) \left[ h(q_C,p_C) \right]_\text{restricted}~,
\end{align}
or in operator form
\begin{align}
    \hat{A} = \hat{A}_Q \left[ {A_C} (q_C,p_C) \right]_\text{restricted}~.
\end{align}

An important consequence of this restriction is that a class of quantum-classical interactions is now ruled out.  Since interaction is represented by a hybrid term coupling quantum and classical variables in the Hamiltonian, only interaction terms that belong to the $\circledast$-associative subalgebra are allowed.  Interestingly, since the $\circledast$-product is different for different quantization schemes on the $C$ sector, it follows that certain interactions would be allowed under one $\circledast$ product while prohibited under another.

\subsection{Examples of allowed hybrid variables}\label{sec:examples}

We now give a few examples of $\circledast$-associative hybrid variables defined by the symmetric product $\sigma$ of the form~\eqref{eq:sigma-abc}.  For any choice of $(a,b,c)$, any linear combination of $q_C$ and $p_C$ can consistently couple to quantum variables.  That is, any hybrid variable of the form
\begin{align}
    f(q_Q,p_Q,q_C,p_C) = g(q_Q,p_Q)\cdot(\alpha q_C + \beta p_C)
\end{align}
is $\circledast$-associative.

Another example is for any bracket with $a=0$.  The classical part of hybrid variables can be any function of $q_C$ only:
\begin{align}
    f(q_Q,p_Q,q_C) &= g(q_Q,p_Q)\cdot h(q_C)~.
\end{align}
We see that this covers a wide class of interesting couplings.  Conversely, for $(a,b=0,c)$, any function of $p_C$ only, or
\begin{align}
	f(q_Q,p_Q,p_C) &= g(q_Q,p_Q)\cdot h(p_C)~,
\end{align}
can couple to quantum variables.

A less obvious example is found for the choice $(a=0,b=0,c=\pm1)$ (obtained from the standard and antistandard operator orderings).  For this choice, any hybrid variable of the form
\begin{align}
    f(q_Q,p_Q,q_C,p_C) = g(q_Q,p_Q)\cdot(\alpha q_C + \beta p_C + \gamma q_Cp_C)
\end{align}
will have consistent hybrid dynamics.

\subsection{Minimal subalgebra}\label{sec:minimal}

One simplifying direction is to consider a class of functions that reduce the semiclassical $\circledast$-product [Eq.~\eqref{eq:ast-pc}] to ordinary multiplication:
\begin{align}\label{eq:minimal}
    f_1(q_C,p_C) \circledast f_2(q_C,p_C) = f_1(q_C,p_C) \cdot f_2(q_C,p_C)~.
\end{align}
Ordinary multiplication is of course associative.

The hybrid composition product $\circledast$ is necessary for consistent hybrid dynamics, as shown in the preceeding section; however, the interpretation of composing classical variables using that product is unclear.  The condition~\eqref{eq:minimal} aims to further narrow down the allowed hybrid variables to those for which the composition product has the familiar meaning of ordinary multiplication.  In a sense, the resulting subalgebra is \textit{minimal}.

Let $\kappa(q_C,p_C)$ be a classical variable belonging to the minimal subalgebra.  Then from~\eqref{eq:minimal} and the definition of $\circledast$ [Eq.~\eqref{eq:ast-pc}] we have
\begin{align}
    \kappa \circledast \kappa = \kappa^2 + \frac{i\hbar}{2} \kappa \, \sigma \, \kappa = \kappa^2 \\
    \Rightarrow \kappa \, \sigma \, \kappa = 0~. \label{eq:k-s}
\end{align}
If we consider the form~\eqref{eq:sigma-abc} of $\sigma$ as a symmetric first-order bidifferential operator, then, condition~\eqref{eq:k-s} becomes
\begin{align}\label{eq:k-cond}
    a\left(\frac{\partial\kappa}{\partial q_C}\right)^2+
    b\left(\frac{\partial\kappa}{\partial p_C}\right)^2+
    2\,c\,\frac{\partial\kappa}{\partial q_C}\,\frac{\partial\kappa}{\partial p_C}=0~.
\end{align}

Now that we found one member of the minimal subalgebra, we can generate an infinite class of related members.  It easy to show that if $\kappa$ obeys~\eqref{eq:k-cond}, then any function $f(\kappa)$ also belongs to the minimal subalgebra:
\begin{align}
    f_1(\kappa) \circledast f_2(\kappa) = f_1(\kappa) \cdot f_2(\kappa)~.
\end{align}
Any hybrid variable of the form
\begin{align}
    f(q_Q,p_Q,q_C,p_C) = g(q_Q,p_Q) \cdot h\big( \kappa(q_C,p_C) \big)~,
\end{align}
where $\kappa(q_C,p_C)$ is a solution to~\eqref{eq:k-cond}, belongs to consistent hybrid dynamics.

An example of $\kappa$ is the simple solution
\begin{align}
    \kappa(q_C,p_C) = q_C \pm \frac{\mp c + \sqrt{c^2 - ab}}{b} \, p_C
\end{align}
for all choices of $(a,b,c)$.  A special case is obtained for $(a=1,b=1,c=0)$ (obtained from the Husimi realization),
\begin{align}
    \kappa(q_C,p_C) = q_C \pm i p_C~,
\end{align}
which bears a resemblance to the quantum ladder operators.

Finally, we stress that pure classical variables are totally free to be any function of $q_C$ and $p_C$.  It is the classical factor of \textit{hybrid} variables that needs to be restricted.

Now that we have more hybrid variables admitted into the theory, in the next section we study time evolution defined in terms of the hybrid bracket.


\section{General dynamics}\label{sec:dynamics}

A complete description of quantum systems includes the \textit{state} of the system in addition to the dynamics of observables.  Thus a discussion of the state of a hybrid system is necessary.  This is yet another motivation for carrying out the analysis of quantum-classical systems in the language of the phase-space formulation.

\subsection{Heisenberg picture}\label{sec:hp}

In the Heisenberg picture, the equations of motion evolve dynamical variables (e.g., observables) in time, while the \textit{state} encodes the initial conditions.  In classical mechanics, the state distribution can be a Dirac $\delta$ function, reflecting complete knowledge of the positions and momenta.  It can also be a different distribution, reflecting partial or incomplete knowledge of the system.

Classical expectation values are calculated as
\begin{align}
    \langle f_C \rangle (t) = \int dq_C\,dp_C ~ \rho_C(0) \, f_C (t)~.
\end{align}
Even though state distributions do not evolve in time in the Heisenberg picture, we write $\rho_C(0)$ instead of simply $\rho_C$ to emphasize that they contain information about the initial conditions.  The dynamical variable $f_C$ of course evolves according to its equation of motion $df_C/dt = \{f_C,H\} + \partial f_C/\partial t$~.

Quantum systems can be described by a normalized phase-space state distribution (quasiprobability distribution) $\rho_Q$.  It is a representation of the state operator $\hat{\rho}_Q$.  For example, in the Wigner representation, the state distribution is the Wigner function $\rho_Q^{(W)}$, the Wigner transform of the density matrix $\hat{\rho}_Q$,
\begin{align}
    \rho_Q^{(W)} &= \frac{1}{(2\pi\hbar)^N} \int dq' \text{e}^{ipq'/\hbar}\left\langle q-\frac{1}{2}q'\right|\hat{\rho}_Q\left| q+\frac{1}{2}q'\right\rangle~.
\end{align}
The Wigner distribution is of course one of many possible quantum distributions.  Another class can be obtained through a convolution of a given state distribution to produce a new state distribution.  An example of this is the Husimi distribution which is a special case of a Gaussian smoothing of the Wigner distribution (see \cite{LeeHW14GSmoothedWigner} and references therein; see also \cite{RobbinsWalton18Augmented}).

Expectation values of quantum variables $f_Q$, similar to classical ones, are calculated in terms of the quantum phase-space state distribution
\begin{align}
    \langle f_Q \rangle (t) = \int dq_Q\,dp_Q ~ \rho_Q(0) \, f_Q (t)~.
\end{align}
Quantum dynamical variables evolve according to the phase-space analog of the Heisenberg equation of motion $df_Q/dt = [\![f_Q,H]\!] + \partial f_Q/\partial t$~.

Now consider interacting quantum-classical systems.  As shown in Sec.~\ref{sec:interaction}, the dynamics of pure quantum or classical variables can be defined using the reduction properties~\eqref{eq:reduction} of the hybrid bracket.  The dynamics of hybrid variables, however, require the full definition of the hybrid bracket.  Otherwise, the time derivative of a hybrid variable $f=f_Q\,f_C$ is ambiguous.  We have seen from~\eqref{eq:c-dot} that $\dot{f}_C$ will in general depend on $Q$ variables.  Thus, naively applying the Leibniz rule for the time derivative $\frac{d}{dt}(f_Q\,f_C)$ will be ambiguous in the ordering of $f_Q$ and $\frac{d}{dt}f_C$~.  On the other hand, it has been illustrated in~\cite{amin2021illustration} that the pure canonical relations for pure variables is not preserved by quantum-classical interaction.

A consistent hybrid bracket is thus required.  Specifically, the Leibniz rule is needed for an unambiguous time evolution, and the Jacobi identity is needed to preserve the canonical relations.  Having defined the consistency conditions in terms of the hybrid composition product and its associative subalgebra in Sec.~\ref{sec:subalg}, we can write the equation of motion as
\begin{align}\label{h-dot}
    \frac{df}{dt} = \{\![f,H]\!\} + \frac{\partial f}{\partial t}~.
\end{align}
This equation describes time evolution of hybrid variables $f=f_Q\,f_C$ in the Heisenberg picture.

One can determine the state distributions $\rho_Q(0)$ and $\rho_C(0)$ of the $Q$ and $C$ subsystems \textit{before interaction} using the standard methods of quantum and classical mechanics.  Since initially the two subsystems were not interacting, we hypothesize that the \textit{initial hybrid state distribution} $\rho(0)$ is given by
\begin{align}
    \rho(0) = \rho_Q(0) \, \rho_C(0)~.
\end{align}
The expectation value of a hybrid dynamical variable $f$ would be given by
\begin{align}
    \langle f \rangle (t) = \int dq_Q\,dp_Q \, dq_C\,dp_C ~ \rho(0) \, f (t)~.
\end{align}

\subsection{Schrödinger picture}\label{sec:sp}

Things are not as simple for hybrid dynamics in the Schrödinger picture where dynamical variables are fixed in time and the state distribution evolves.\footnote{For a nice discussion of the equations of motion in the Heisenberg and Schrödinger picture, see~\cite{snygg_heisenberg_1982}.}

In the Schrödinger picture of classical mechanics, the state distribution evolves in time according to the Liouville equation
\begin{align}
    \frac{d\rho_C}{dt} = \{\rho_C,H\} + \frac{\partial\rho_C}{\partial t} = 0~.
\end{align}
Classical expectation values are given by
\begin{align}
    \langle f_C \rangle (t) = \int dq_C\,dp_C ~ \rho_C(t) \, f_C (0)~.
\end{align}
Notice the switch in time dependence between $\rho_C$ and $f_C$~.

The quantum state distribution evolves in time according to the phase-space analog of the von Neumann equation
\begin{align}
    \frac{d\rho_Q}{dt} = [\![\rho_Q,H]\!] + \frac{\partial\rho_Q}{\partial t} = 0~.
\end{align}
Quantum expectation values are given by
\begin{align}
    \langle f_Q \rangle (t) = \int dq_Q\,dp_Q ~ \rho_Q(t) \, f_Q (0)~.
\end{align}

For hybrid variables, things should work the same way with expectation values given by
\begin{align}
	\langle f \rangle (t) = \int dq_Q\,dp_Q \, dq_C\,dp_C ~ \rho(t) \, f (0)~,
\end{align}
except that now defining the hybrid state distribution $\rho$ is not straightforward, since the two subsystems are interacting.  The hybrid equation of motion for the state distribution
\begin{align}
\frac{d\rho}{dt} = \{\![\rho,H]\!\} + \frac{\partial\rho}{\partial t} = 0
\end{align}
involves the hybrid bracket of two hybrid variables.

As discussed in Sec.~\ref{sec:subalg}, for the bracket to be consistent, $\rho$ and $H$ must belong to the same $\circledast$-associative subalgebra.  While the equivalence between the Heisenberg and Schrödinger pictures in classical and quantum mechanics is established, it is unclear at this stage under what conditions this equivalence is true in hybrid dynamics.  We leave this problem for future work.


\section{Conclusion}\label{sec:summary}

Using the phase-space formulation of quantum mechanics, a general hybrid quantum-classical bracket~\eqref{eq:taste2} [or~\eqref{eq:hybrid1}, \eqref{eq:hybrid2} or~\eqref{eq:hybrid3}] was easily derived from a partial classical limit. In the Wigner-Weyl-Moyal representation, it coincides with the hybrid bracket~\eqref{eq:AGBT} proposed in \cite{aleksandrov_statistical_1981, gerasimenko_1982, boucher_semiclassical_1988}, but differs in other realizations.  The bracket takes the form of the commutator of a hybrid composition product and thus obeys the Jacobi identity and Leibniz rule if that product is associative.

Pure quantum and classical variables can consistently interact with each other through a hybrid Hamiltonian.  In general, if only one entry of the hybrid bracket is hybrid, then consistency (the Jacobi identity and the Leibniz rule, in particular) is guaranteed.  Thus, one can choose to work in the Heisenberg picture where the state distribution does not evolve in time, and have the Hamiltonian as the only hybrid variable in the system.  In that case, interaction between quantum and classical variables, and quantum backreaction can be meaningfully studied using the hybrid bracket.

When one is interested in studying the dynamics of hybrid variables rather than interacting pure quantum and classical ones, then restrictions apply.  Composition products derived from the partial classical limit are not generally associative, yet they define associative subalgebras of hybrid variables.  The class of hybrid variables that belong to an associative subalgebra then has consistent dynamics.  Restricting hybrid variables to those belonging to an associative subalgebra with the hybrid composition product will necessarily restrict interaction Hamiltonians.

In the Schrödinger picture, the state distribution is a hybrid variable that evolves in time (through a bracket with the hybrid Hamiltonian).  The possibility of constructing suitable quantum-classical state distributions that belong to the allowed set of hybrid variables is still under investigation.  We see that the consistency of quantum-classical dynamics restricts possible quantum-classical interactions and states.

\begin{acknowledgments}
	This research was supported in part by a Discovery Grant (M.A.W.) from the Natural Sciences and Engineering Research Council (NSERC) of Canada (funding reference No. RGPIN-2015-05809).
\end{acknowledgments}


\bibliography{refs}

\end{document}